\documentclass[12pt, preprint]{aastex}

\usepackage{graphicx}
\usepackage{amssymb}
\usepackage{layout}
\usepackage{url}
\usepackage{subfigure}
\usepackage{color}
\slugcomment{Submitted to {\it The Astrophysical Journal}}

\shorttitle{RMHD Shock in Helical Magnetic Field in Blazars}
\shortauthors{H. Zhang, W. Deng, H. Li, \& M. B\"ottcher}

\begin{document}


\title{Polarization Signatures of Relativistic Magnetohydrodynamic Shocks in the Blazar Emission Region - 
I. Force-free Helical Magnetic Fields}


\author{Haocheng Zhang\altaffilmark{1,2}, Wei Deng\altaffilmark{3,2}, Hui Li\altaffilmark{2} and 
Markus B\"ottcher\altaffilmark{4}}

\altaffiltext{1}{Astrophysical Institute, Department of Physics and Astronomy, \\
Ohio University, Athens, OH 45701, USA}

\altaffiltext{2}{Theoretical Division, Los Alamos National Laboratory, Los Alamos, NM 87545, USA}

\altaffiltext{3}{Department of Physics and Astronomy, University of Nevada Las Vegas, Las Vegas, NV 89154, USA}

\altaffiltext{4}{Centre for Space Research, North-West University, Potchefstroom, 2520, South Africa}

\begin{abstract}
The optical radiation and polarization signatures in blazars are known to be highly variable 
during flaring activities. It is frequently argued that shocks are the main driver of the flaring 
events. However, the spectral variability modelings generally lack detailed considerations of the 
self-consistent magnetic field evolution modeling, thus so far the associated optical polarization 
signatures are poorly understood. We present the first simultaneous modeling of the optical
radiation and polarization signatures based on 3D magnetohydrodynamic simulations of relativistic 
shocks in the blazar emission environment, with the simplest physical assumptions. By comparing the 
results with observations, we find that shocks in a weakly magnetized environment will largely lead 
to significant changes in the optical polarization signatures, which are seldom seen in observations. 
Hence an emission region with relatively strong magnetization is preferred. In such an environment, 
slow shocks may produce minor flares with either erratic polarization fluctuations or considerable 
polarization variations, depending on the parameters; fast shocks can produce major flares with 
smooth PA rotations. In addition, the magnetic fields in both cases are observed to actively 
revert to the original topology after the shocks. All these features are consistent with 
observations. Future observations of the radiation and polarization signatures will further 
constrain the flaring mechanism and the blazar emission environment.
\end{abstract}

\section{Introduction}

Blazars are the most violent active galactic nuclei. They are known to emit nonthermal-dominated 
radiation from radio to $\gamma$-rays with strong variability across the entire electromagnetic 
spectrum. It is generally agreed that the emission comes from an unresolved region in a relativistic 
jet that is directed close to our line of sight (LOS). The blazar spectrum has two components. The 
low-energy component, from radio to optical/UV, is known to be polarized, with the polarization 
percentage ranging from a few to tens of percent. This is in agreement with the synchrotron 
emission from nonthermal electrons in a partially ordered magnetic field. Several papers have 
demonstrated that the observed polarization signatures may indicate a helical magnetic structure 
\citep{Lyutikov05,Pushkarev05,Zhang15}. The high-energy component, from X-rays to $\gamma$-rays, 
is usually interpreted as originating from inverse Compton scattering by the same nonthermal 
electrons of soft seed photons \citep[e.g.,][]{Marscher85,Dermer92,Sikora94}, but a hadronic 
origin cannot be ruled out \citep[e.g.,][]{Mannheim92,Mucke01,Boettcher13}. Both spectral 
components of blazars exhibit fast variability. Many observations show flares in various 
observational bands lasting days or even hours \citep[e.g.,][]{Ciprini11,Chatterjee12}; in particular, 
high-energy $\gamma$-rays sometimes show variability time scale within several tens of minutes 
\citep[e.g.,][]{Aharonian07,Albert07}. Based on the causality relation, $R\lesssim \delta c t$, 
where $\delta$ is the Doppler factor and $t$ is the observed flaring time scale, the size of the 
emission region $R$ in the comoving frame of the emission region should generally be within 
$0.1~pc$. Therefore, it is often argued in the blazar spectral variability fittings that the 
blazar emission region is a small region near the broad line region of the blazar jet 
\citep[e.g.,][]{Tavecchio10,Boettcher13,Barnacka14}.

Blazar spectral fittings have shown that the nonthermal electron spectra responsible for the common 
blazar SEDs require power-law indices of $\gtrsim 2$ in most cases \citep[e.g.,][]{Boettcher13}. 
This is in agreement with numerical simulations of relativistic shocks, which have 
demonstrated that diffusive shock acceleration forms such power-law spectra 
\citep{Achterberg01,Spitkovsky08,Summerlin12}. Therefore, shocks are frequently 
used to explain blazar flaring activities. However, in general, numerical simulations 
of shock acceleration do not provide detailed calculations of the expected radiation 
features. In terms of the spectral variability fittings, scenarios of relativistic shocks 
propagating through the jet have been widely investigated
\citep[e.g.,][]{Marscher85,Spada01,Joshi07,Graff08}. These models have successfully fit 
the time-dependent SEDs and multiwavelength light curves of blazar flares, but they usually 
assume a stationary chaotic magnetic field. Therefore, the polarization variations are poorly 
understood and/or constrained.

Polarimetry plays an important role in constraining the jet physics. Radio polarimetry 
has been a standard tool to understand magnetic fields of large-scale jets in radio galaxies 
\citep[e.g.,][]{Laing14}. Additionally, sometimes blazar flaring activities 
have been reported to be connected with changes in radio knots \citep[e.g.,][]{Marscher08}. 
However, since blazar emission regions remain spatially unresolved, high-resolution imaging
and polarimetry at radio wavelengths usually cannot directly constrain the physical 
conditions in the blazar high-energy emission environment. However, the optical emission 
from the blazar emission region generally dominates 
over all the other optical emissions from the blazar jet, hence optical polarimetry is able 
to directly reveal the inner-jet magnetic field structure. The observed optical polarization 
signatures are generally erratic. Typically, the polarization degree (PD) varies within 
$\sim 30\%$, but occasionally a higher PD ($\sim 40\%$) is reported \citep[e.g.,][]{Scarpa97}. 
The polarization angle (PA) usually displays perturbations around some mean values. However, 
polarization signatures can be highly variable during the flaring activities. In particular, 
significant multiwavelength flaring activities are seen to be accompanied by large 
($\gtrsim 180^{\circ}$) PA rotations \citep[e.g.,][]{Marscher08,Abdo10,Chandra15}. 
These flare + PA rotation events often feature smooth PA rotations and apparently 
time-symmetric light curves and PD patterns. In addition to the individual observations, 
\cite{Blinov15} are performing a polarization monitoring program of a large sample, 
in which they find that blazars with detected PA rotations generally exhibit stronger 
variations in the PA, and the rotations can go in both directions. Moreover, they claim 
that the PA rotations are probably physically connected to the $\gamma$-ray flares.

In terms of modeling, several mechanisms have been put forward to understand the polarization 
signatures, especially the PA rotations. An emission region following a helical trajectory 
\citep[e.g.,][]{Marscher08} can give rise to a smooth PA rotation, but such a model prefers 
all rotations present in the same blazar to follow the same rotating direction, which 
contradicts observations. An initially chaotic magnetic field structure compressed by 
a flat shock \citep{Laing80} or multiple small shocks \citep{Marscher14} can produce 
erratic polarization patterns and occasionally large polarization variations through 
random walks, but the resulting patterns are normally whimsical, and it has been shown 
that the observed PA rotations are unlikely to result completely from random walks 
\citep{Kiehlmann13,Blinov15}. \cite{Chen14} and \cite{Zhang14} have proposed a model 
in which a disturbance propagates through a cylindrical emission region pervaded by 
a helical magnetic field. This model can give rise to systematic PA rotations and 
apparently time-symmetric light curves and PD patterns. Based on this model, \cite{Zhang15} 
and \cite{Chandra15} have presented simultaneous fittings of SEDs, light curves and 
polarization signatures for two flare + PA rotation events. Nonetheless, in all the 
models mentioned above, the magnetic field evolution is treated in an ad hoc way. 
Furthermore, none of them can so far explain the statistical properties of the 
polarization variations as shown in \cite{Blinov15}.

In this paper, we present the first polarization-dependent radiation modeling based 
on relativistic magnetohydrodynamics (RMHD) simulations of shocks in helical 
magnetic fields in the blazar emission region. This new approach performs 
detailed analysis on the interaction between the shock propagation and the magnetic 
field evolution, which enables us to constrain the blazar emission environment and 
the shock parameters. We demonstrate that the blazar emission region largely possesses 
significant magnetic energy. In such an environment, fast shocks can be the driver of 
flaring events, producing both the typical polarization fluctuations and the occasional 
large PA rotations. Our simulation results are consistent with the statistical polarization 
properties reported in \cite{Blinov15}. We will describe our model setup in Section 2, 
illustrate the interaction between the shock and the magnetic field as well as the 
consequent radiation and polarization signatures in Section 3, present additional 
parameter studies in Section 4, and discuss the results in Section 5.

\section{Model Setup}

The purpose of this paper is to study the time-dependent radiation and polarization 
signatures resulting from an RMHD shock in the blazar emission environment, with the simplest 
physical assumptions. In this section, we will describe our physical assumptions and 
corresponding code structure in detail.

\subsection{Physical Assumptions}

The blazar emission region is often considered to be an unresolved region near the broad 
line region of the jet. Generally speaking, the magnetic field topology and evolution as well 
as plasma flow dynamics in the blazar emission region should be linked with those in the 
large-scale jet. However, the oberved fast variability and high luminosity suggest 
that the emission region is an extraordinary, very small and localized region with fast evolution. Therefore, 
we argue that within the time scale of an individual flare, the blazar emission region can be 
considered as uncorrelated with the properties of the large-scale jet, even though it is 
spatially embedded within the jet. On the other hand, the emission region is expected to
remain well localized in the jet during flares due to pressure provided by the large-scale 
jet structure. We assume that flares are due to a disturbance propagating through the emission 
region. Many observations show that light curves and even the time-dependent polarization 
signatures appear symmetric in time \citep[e.g.,][]{Abdo10}. This means that the emission 
region and the disturbance are likely to possess some kind of symmetry. \cite{Zhang14,Zhang15} 
have shown that a helically symmetric emission region and disturbance, with detailed 
consideration of light-travel-time effects (LTTEs), can naturally explain the apparently 
time-symmetric light curves and time-dependent polarization signatures. Therefore, we 
continue to use that assumption here.

Due to the relativistic aberration, even though we are observing blazars nearly along the 
jet in the observer's frame (typically, $\theta_{obs,1} \sim 1/\Gamma_1$), where $\theta_{obs,1}$ 
and $\Gamma_1$ are the angle between the LOS and the jet direction and the Lorentz factor of 
the emission region in the observer's frame, respectively, the angle $\theta_{obs}$ between 
LOS and the jet axis in the comoving frame is likely around $90^{\circ}$ (if $\theta_{obs,1} 
\sim 1/\Gamma_1$, then $\theta_{obs} \sim 90^{\circ}$). \cite{Zhang14} have shown that if the LOS is fixed at some other angles, the general trend of polarization variations is only weakly affected. In addition, as is suggested in the bending jet scenario of the polarization signatures, a change in the LOS direction may significantly affect the polarization signatures. However, those effects are beyond the scope of our first-step study. Thus in all of the following simulations, we choose $\theta_{obs} = 90^{\circ}$ in the comoving frame, and hence the Doppler factor $\delta \equiv \left( \Gamma_1 \, [1 - \beta_{\Gamma_1} \, \cos\theta_{obs,1}] \right)^{-1} \sim \Gamma_1$.

We use ideal RMHD simulations to describe the evolution of the system. The underlying model 
assumes that a plasma jet is traveling with a Lorentz factor of $\Gamma_0$ in the observer's 
frame. In the comoving frame of the jet, a flow of plasma which is pervaded by a helical 
magnetic field travels at a Lorentz factor of $\Gamma$. Inside this flow lies a cylindrical 
emission region containing nonthermal particles. Along the path of the flow it encounters a 
flat stationary layer of the plasma (the disturbance), forming a shock wave. Thus the total 
Lorentz factor of the emission region in the observer's frame is $\Gamma_1=\Gamma\times\Gamma_0$. 
Although in the following sections our RMHD simulations use different Lorentz factors $\Gamma$ 
of the disturbance in the comoving frame of the emission region, $\Gamma_1$ of the emission 
region in the observer's frame is not constrained. Therefore, we assume a fixed Lorentz factor 
of $\Gamma_1=20$ for the emission region in the observer's frame, which is commonly inferred 
from SED modelings.

Before the emission region interacts with the disturbance, the helical magnetic field is 
assumed to be in force-balance. The advantage of this magnetic field setup is 
that it will naturally give rise to comparable poloidal and toroidal contributions, resulting 
in a relatively low background PD without the need of any turbulence. In addition, this can 
simplify our setup for the plasma density and the thermal pressure, which are taken to be 
uniform inside the emission region. In this way, we assume that initially all flow conditions 
and magnetic fields are laminar, with no preexisting turbulent components. Also, as the RMHD 
requires an input for the thermal pressure, we assume that in the beginning the plasma is cold 
and hence the thermal pressure is very small compared to the kinetic energy or the magnetic 
energy.

In the comoving frame of the emission region, the disturbance and the resulting shock will propagate 
through the emission region, and change the physical conditions as well as accelerate particles at 
its location, generating a flare. We apply the simple assumption that the shock will inject fresh 
nonthermal electrons at the shock front whose energy is equal to a fixed small amount of the local 
shock kinetic energy. In this way, any deceleration and deformation of the shock due to the magnetic 
field obstruction can be taken into consideration. In Section 4 we will show that the injection rate 
does not affect the polarization signatures too much.

We briefly summarize our model assumptions in the following:\\
1. The emission region is a small cylindrical region embedded in the large-scale jet.\\
2. The evolution of the emission region is detached from the large-scale jet.\\
3. The boundary of the emission region is held by a pressure wall, probably provided by the 
large-scale jet.\\
4. In the comoving frame of the emission region, the observer is observing from the side of 
the emission region ($\theta_{\rm obs} = 90^o$).\\
5. The initial magnetic field inside the emission region is a force-free helical magnetic field.\\
6. Initially all flow conditions and magnetic fields are laminar.\\
7. Initially the plasma is cold.\\
8. Initially the disturbance is a flat cylindrical region traveling relativistically in the comoving 
frame of the emission region.\\
9. The disturbance will generate a shock in the emission region. At the shock front, the shock will 
inject nonthermal particles whose total energy is a constant, small fraction of the local shock 
kinetic energy.

\begin{figure}[ht]
\centering
\includegraphics[width=0.9\linewidth]{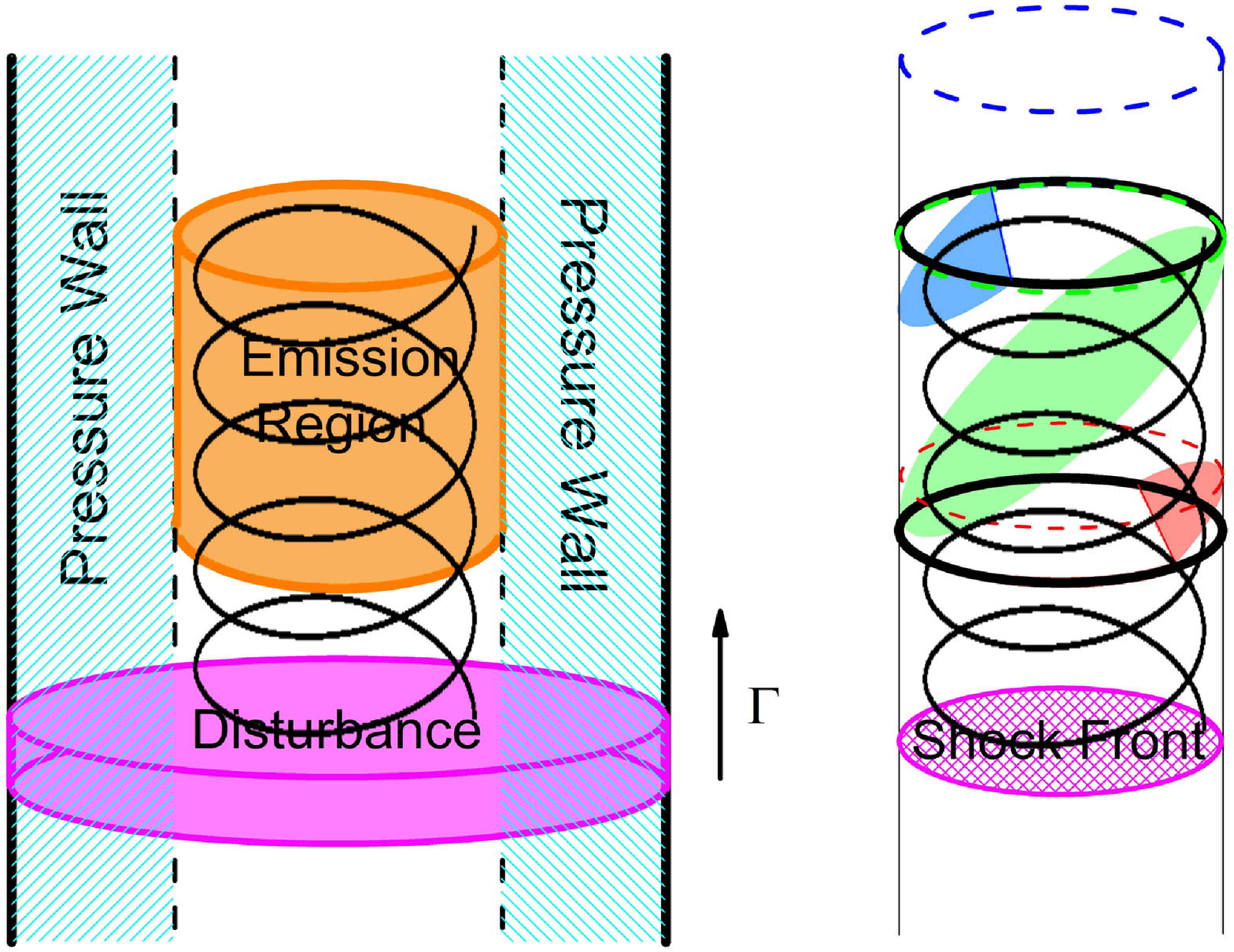}
\caption{Left: a sketch of the model setup. The cylindrical simulation domain is pervaded by a 
helical magnetic field with exponential cutoff at the edge, which is held by a pressure wall. 
The emission region has a fixed height, while its radius extends to roughly the pressure wall 
where the magnetic field has been exponentially cut off. The disturbance is a flat region 
traveling upward in the simulation frame. It will form a shock wave which will propagate 
through the emission region and modify the local plasma conditions, and inject fresh nonthermal 
particles at its front. As the disturbance and the shock are relativistic, they are likely to 
stick together during the propagation. Right: a sketch of the LTTEs. The shock/disturbance will 
propagate through the emission region in the comoving frame of the emission region. The red, 
green and blue dash circles refer to the location of the shock front at approximately the 
rising, peak and declining phases of the flare, respectively; the corresponding shapes and 
locations of the flaring region, indicating points of equal photon arrival times at the 
observer, are shown by the light red, light green and light blue shapes, 
respectively. Notice that in our simulations, the shock front is not necessarily flat as in 
the sketch. \label{sketch}}
\end{figure}

\subsection{Code Structure}

The above model is realized by the combination of the 3D multi-zone RMHD code LA-COMPASS 
developed by \cite{Li03} and the 3D multi-zone polarization-dependent ray-tracing code 3DPol 
developed by \cite{Zhang14}. Fig. \ref{sketch} (left) shows a sketch of the simulation setup, 
and Table \ref{parameters} presents some major parameters. The RMHD simulation is performed 
in the comoving frame of the emission region, using Cartesian coordinates. We apply the outflow 
boundary conditions for the simulation domain; in the following, we will demonstrate that our 
setup should not suffer from any boundary effects. The simulation domain is pervaded by a 
helical magnetic field (Fig. \ref{sketch} (left)). In reality, the poloidal magnetic field 
lines have to be closed, but based on our assumption the returning magnetic flux is generally 
outside the simulation domain. The helical magnetic field is assumed to be in force-balance, in the form of
\begin{equation}
\begin{array}{c}
B_z = B_0\times J_0(kr) \\
B_{\phi} = B_0\times J_1(kr)
\end{array}
\end{equation}
where $J_0$ and $J_1$ are the Bessel functions of the first kind, $r$ is the radius, and $B_0$ 
and $k$ are normalization factors of the magnetic field strength and the radius, respectively. 
The $B_r$ component is assumed to be zero. In order to avoid any boundary effects in $x$, $y$ 
directions and comply with the assumed generally monodirectional $B_z$, we add an exponential 
cutoff for both components when $B_z$ approaches zero ($kr\sim 2.4$). However, the exponential 
cutoff will lead to a nontrivial magnetic force at the edge of the simulation domain, hence we 
add a plasma pressure to balance it (Fig. \ref{sketch} (left)).

\begin{table}[ht]
\scriptsize
\parbox{\linewidth}{
\centering
\begin{tabular}{|l|c|c|c|}\hline
Parameters     & Length               & Time               & Velocity                  \\ \hline
Relation       & $L_0$                & $L_0/c$            & $c$                      \\ \hline
Code Unit      & $1$                  & $1$                & $1$                      \\ \hline
Physical Value & $8.33\times10^{16} cm$ & $2.78\times10^6 s$   & $3\times10^{10} cm~s^{-1}$    \\ \hline
Parameters     & Magnetic Field       & Thermal Pressure   & Plasma Density            \\ \hline
Relation       & $B_0$                & $B_0^2/(4\pi)$     & $B_0^2/(4\pi c^2)$       \\ \hline
Code Unit      & $1$                  & $1$                & $1$                      \\ \hline
Physical Value & $0.1 G$              & $8.0\times10^{-4} erg~cm^{-3}$   & $8.8\times10^{-25} g~cm^{-3}$ \\ \hline
\multicolumn{4}{c}{}\\
\end{tabular}}
\parbox{\linewidth}{
\centering
\begin{tabular}{|l|c|}\hline
Bulk Lorentz factor $\Gamma_1$                 & $20$     \\ \hline
Orientation of LOS $\theta_{obs}$ $(^{\circ})$ & $90$     \\ \hline
Electron minimal energy $\gamma_1$             & $10^3$   \\ \hline
Electron maximal energy $\gamma_2$             & $5\times10^4$ \\ \hline
Electron power-law index $p$                   & $2$      \\ \hline
\end{tabular}}
\caption{Summary of parameters. Top: Conversion between the RMHD code units and the physical value. 
Bottom: Additional parameters used in the 3DPol simulation. All parameters in this table are in the 
comoving frame of the emission region, except the bulk Lorentz factor. \label{parameters}}
\end{table}

\begin{figure}[ht]
\centering
\includegraphics[width=\linewidth]{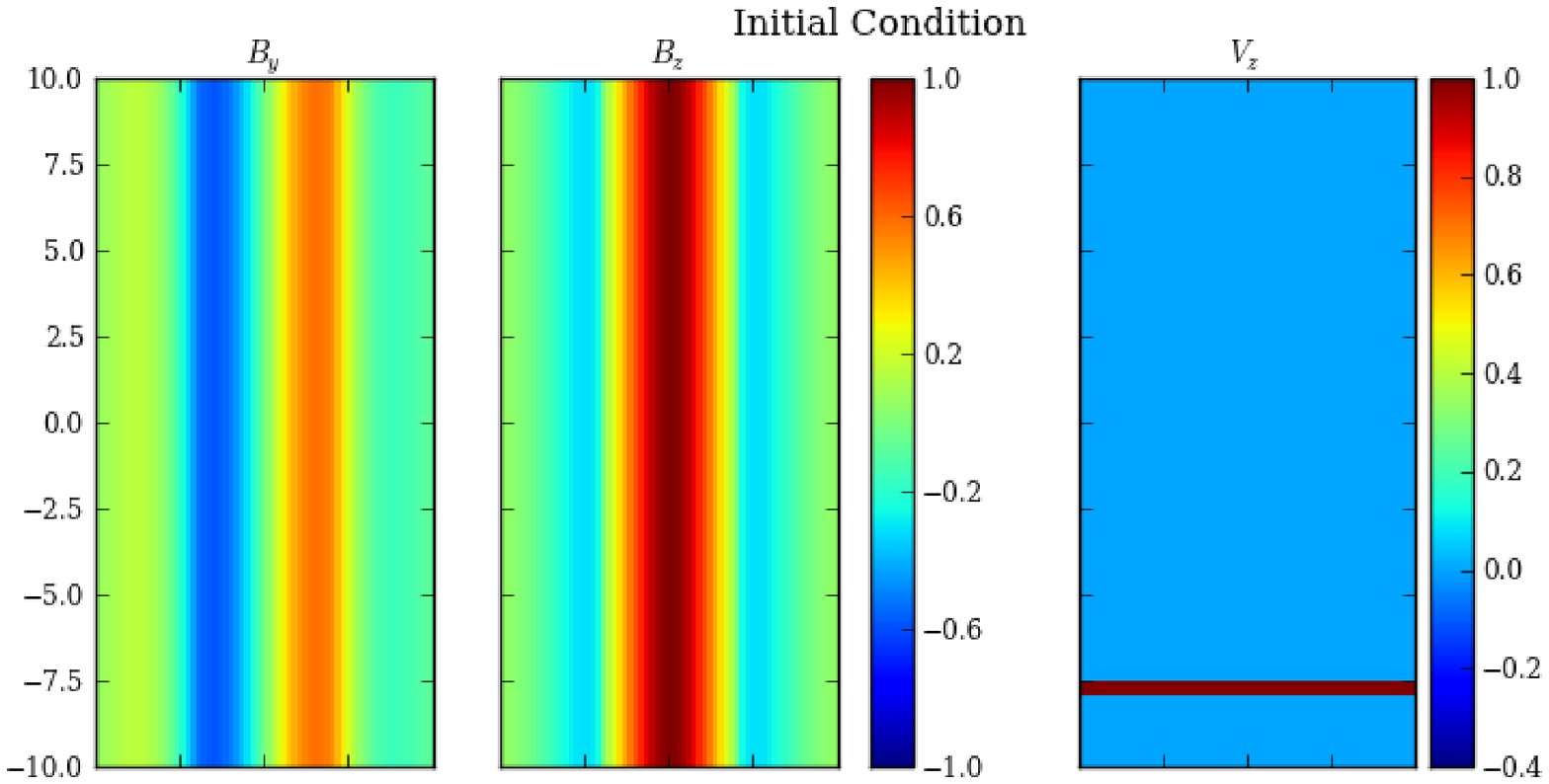}
\caption{Initial condition of the RMHD simulation in the $xz$ plane. We only cut part of 
the simulation in $z$ direction that is related to the emission region. The emission region 
ranges from $-2.5$ to $2.5$. The three panels are the toroidal component $B_y$, which 
represents the $B_{\phi}$ in this plane for a helical magnetic field ($B_x$ is trivial 
in this plane), the poloidal component $B_z$ and the velocity of the disturbance $V_z$. 
All color bars show values in the RMHD code units. \label{mhd0}}
\end{figure}

The emission region is set to be a fixed volume in the simulation domain ($r\sim 2.5$ and $z$ 
ranges from $-2.5$ to $2.5$, see Fig. \ref{mhd0}). Since the simulation domain is in the comoving 
frame of the emission region, the disturbance will travel at a Lorentz factor of $\Gamma$ in the 
$z$ direction. We assume that the disturbance is a thin layer ($0.5$ in width) and initially put 
it at some distance away from the emission region (the disturbance layer ranges from $-8$ to 
$-7.5$) to allow some time for the shock wave to form in the numerical simulation 
(see Fig. \ref{sketch} (left) and Fig. \ref{mhd0}). In order to avoid boundary 
effects in $z$ direction, we use a sufficiently large range in $z$ ($-20$ to $20$), 
so that at the end of the simulation, no signal has reached the $z$ boundary. We 
define a magnetization factor in the emission region,
\begin{equation}
\sigma = \frac{E_{em}}{h}
\end{equation}
where $E_{em}=\frac{B^2+E^2}{8\pi}$ is the electromagnetic energy density, 
$h=\rho c^2+\frac{\hat{\gamma} p}{\hat{\gamma}-1}$ is the specific enthalpy, 
$\rho$ is the plasma density, $\hat{\gamma}$ is the adiabatic index and $p$ 
is the thermal pressure. We further assume that the plasma is cold. Hence we 
choose $\frac{\hat{\gamma} p}{\hat{\gamma}-1}$ to be a small part ($1/30$) of 
the electromagnetic energy initially. In this way the magnetization factor is 
approximately  $\sigma \sim \frac{E_{em}}{\rho c^2}$. The relativistic Alfv\'en 
speed can then be expressed as
\begin{equation}
V_A \sim \frac{c}{\sqrt{4\pi \rho c^2/B^2+1}}
\end{equation}
Since the entire simulation domain is axisymmetric, we present a cut in the $xz$ 
plane in Fig. \ref{mhd0} to illustrate the initial condition, including the magnetic 
field and the velocity of the disturbance. Notice that the $B_x$ component is not 
shown, as in a helical setup, in the $xz$ plane this component is trivial. Also this 
figure is for an initial $\Gamma=10$; for other $\Gamma$ the velocity appears slightly 
different. As the RMHD simulation takes Cartesian coordinates, we will transform the 
results to the cylindrical coordinates and feed into the 3DPol code.

The 3DPol code is focused on the polarization-dependent synchrotron radiation. It uses 
a cylindrical geometry, and further divides it evenly into multiple zones in $r$, $\phi$ 
and $z$ directions. Each zone has its own magnetic field evolution, which is provided by 
the RMHD simulations. The nonthermal electrons in each zone are assumed to range  between 
a minimal and a maximal Lorentz factor $\gamma_1$ and $\gamma_2$, with a fixed spectrum 
of power-law index of $-2$ plus an exponential cutoff at the high-energy end,
\begin{equation}
n(\gamma)=n\gamma^{-2}e^{-\frac{\gamma}{\gamma_2}}, ~ for ~ \gamma>\gamma_1
\end{equation}
The nonthermal electron normalization factor $n$ is assumed to have two components: a 
fixed uniform initial background density $n_b$, and an injected density $n_i$ during 
the shock. As the electron spectrum is fixed, we will only consider the optical light 
curves and polarization patterns, which do not suffer from any synchrotron-self absorption 
and Faraday rotation effects for our parameters. As in the RMHD simulation, all calculations 
are performed in the comoving frame of the emission region. Given the local magnetic field 
information and the nonthermal electron population, the 3DPol code will calculate the Stokes 
parameters at various frequencies at every time step in each zone. Next the code will 
employ ray-tracing to take account of all LTTEs (see Fig. \ref{sketch} (right) for an 
illustration), and add up the incoherent Stokes parameters arriving at the observer at 
the same time, then Lorentz transform to the observer's frame so as to obtain the total 
time-dependent radiation and polarization signatures.

We define the PA in the following way. When the electric vector is parallel to the emission 
region propagation, $PA=0$ (toroidal component dominating). Since the PA has $180^{\circ}$ 
ambiguity, the toroidal dominance happens at $PA=2N\times90^{\circ}$, and the poloidal 
dominance at $PA=(2N+1)\times90^{\circ}$, where $N$ is an integer.

\section{Interaction between Shock and Magnetic Field}

\begin{figure}[ht]
\centering
\includegraphics[width=\linewidth]{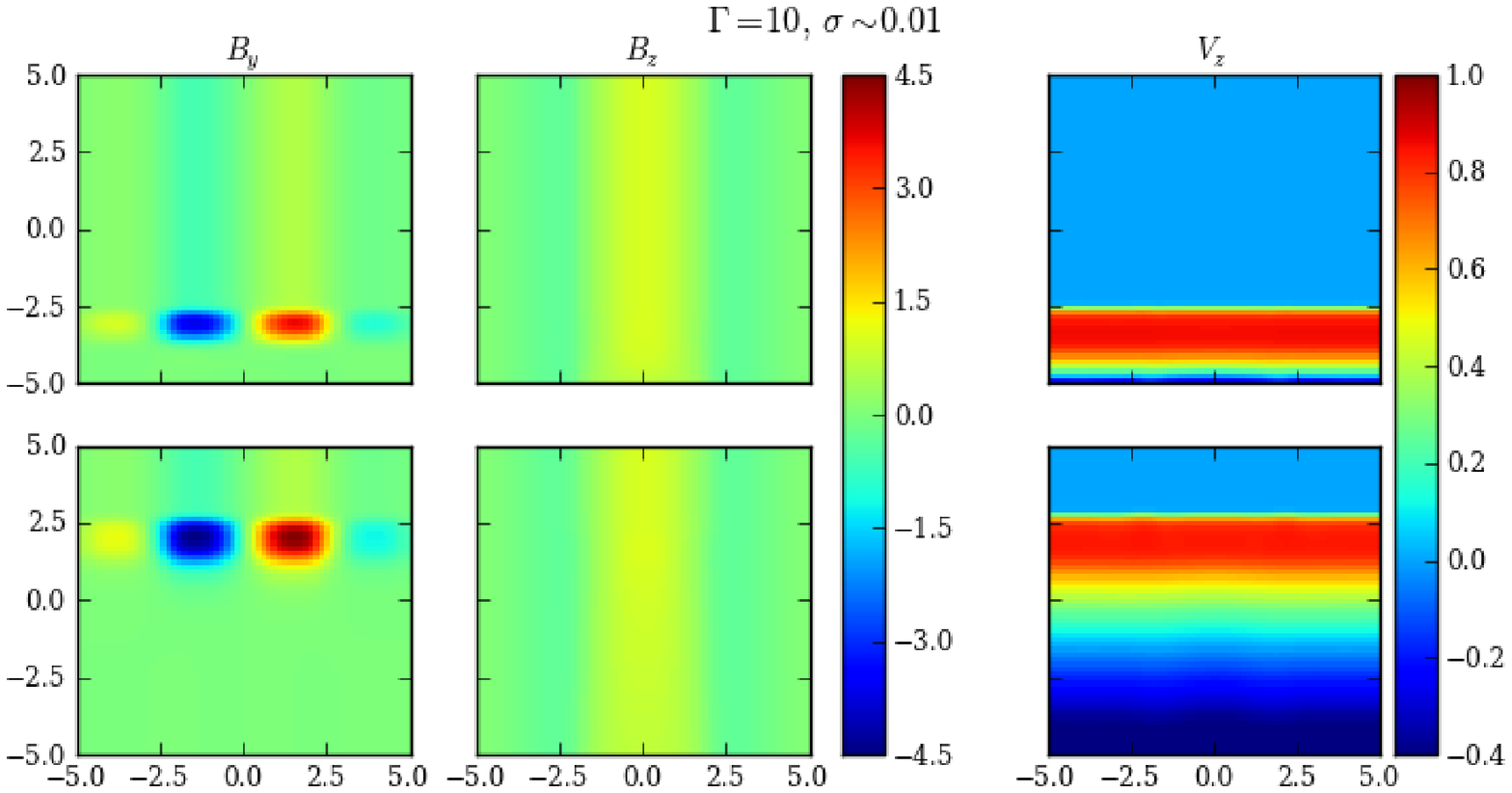}
\caption{RMHD simulation results for Case I. The first row shows the magnetic field 
structures and the plasma velocity when the disturbance is moving into the emission 
region. The second row shows the same information when the disturbance is moving out. 
Notice the color bars for the magnetic fields are different from Fig. \ref{mhd0}, and 
the region is cut smaller. Otherwise plots are as in Fig. \ref{mhd0}. \label{mhd1}}
\end{figure}

\begin{figure}[ht]
\centering
\includegraphics[width=\linewidth]{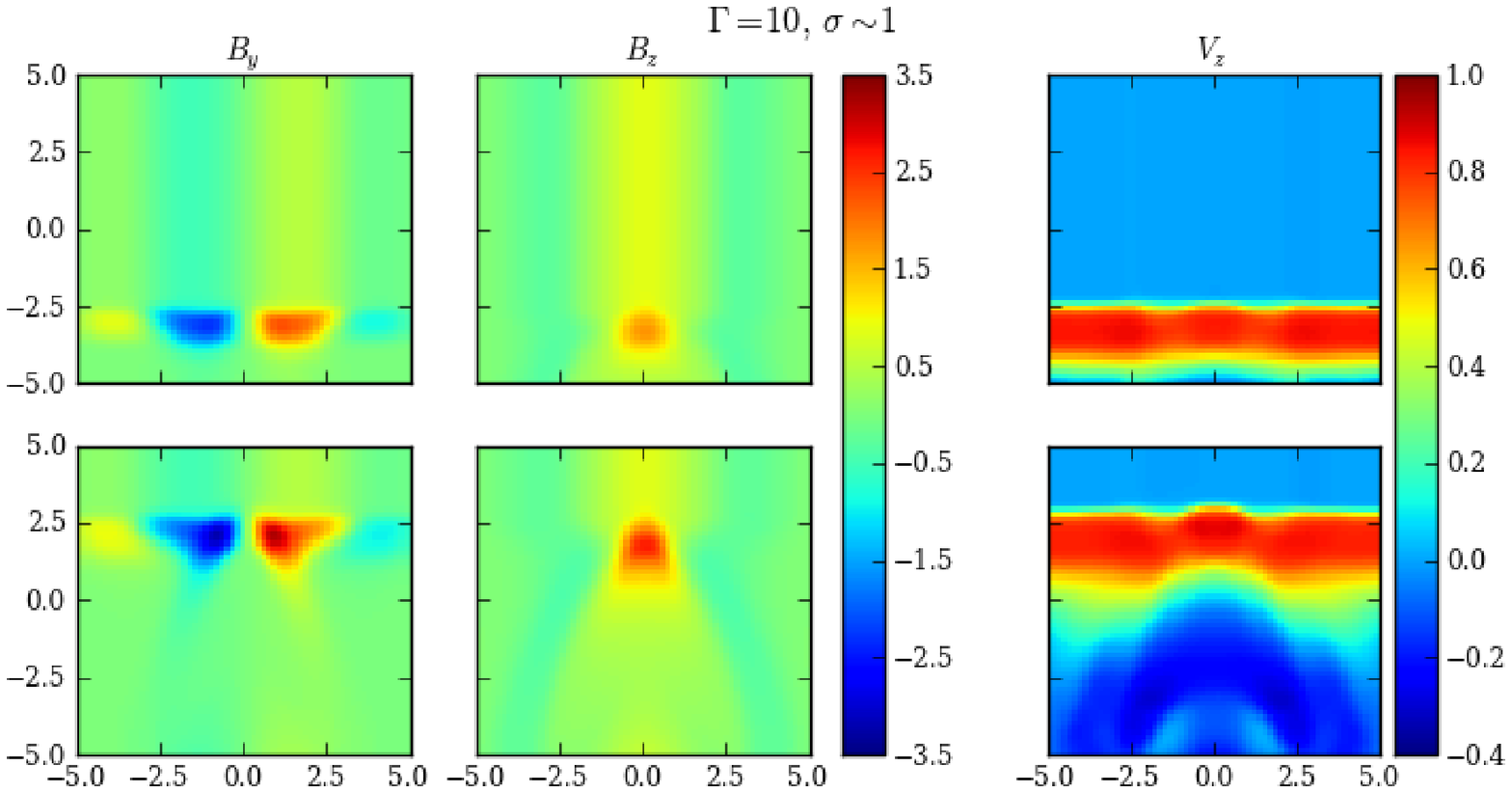}
\caption{RMHD simulation results for Case II. Plots are as in Fig. \ref{mhd1}. \label{mhd2}}
\end{figure}

\begin{figure}[ht]
\centering
\includegraphics[width=\linewidth]{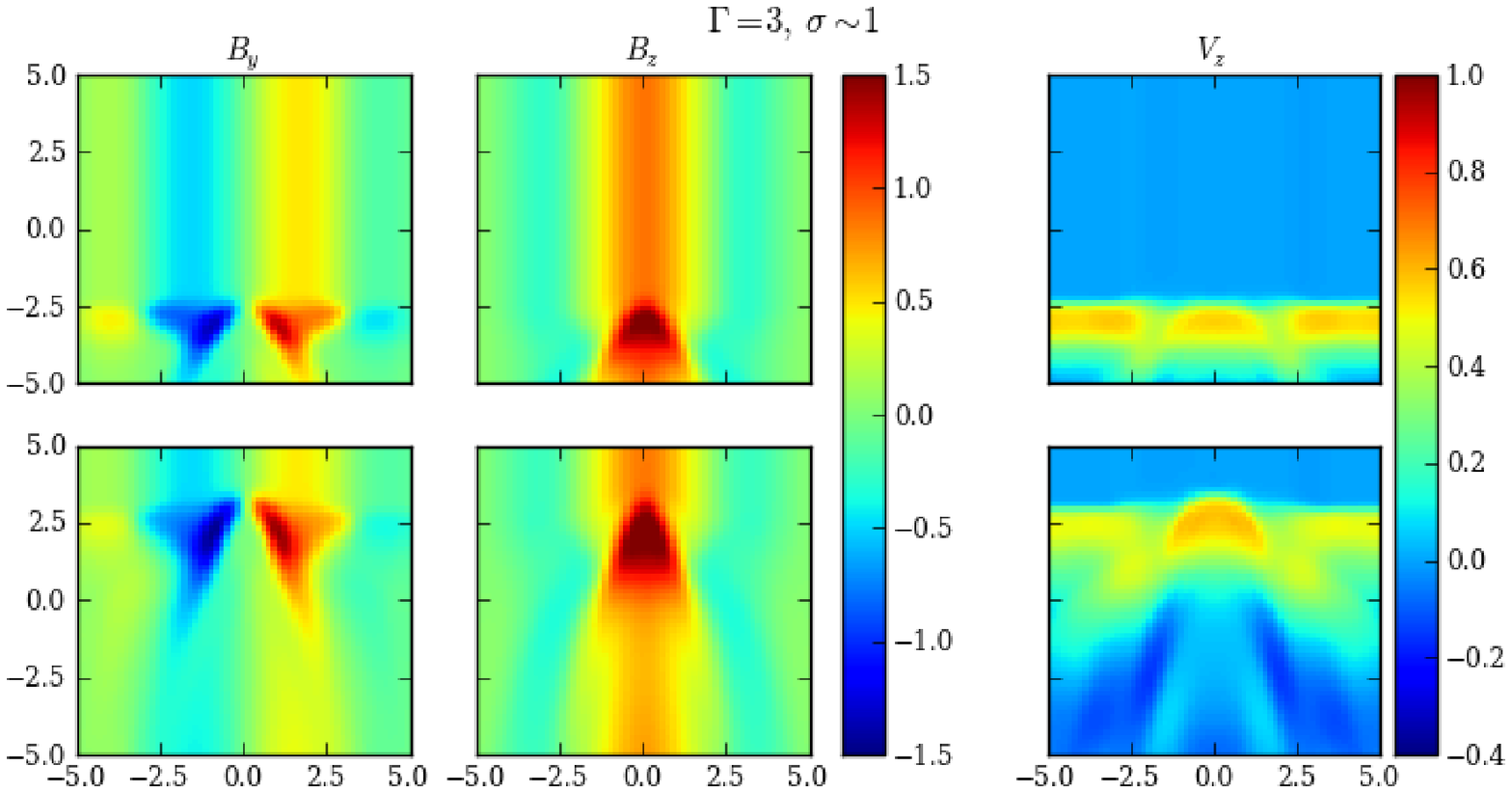}
\caption{RMHD simulation results for Case III. Plots are as in Fig. \ref{mhd1}. \label{mhd3}}
\end{figure}

\begin{figure}[ht]
\centering
\includegraphics[width=\linewidth]{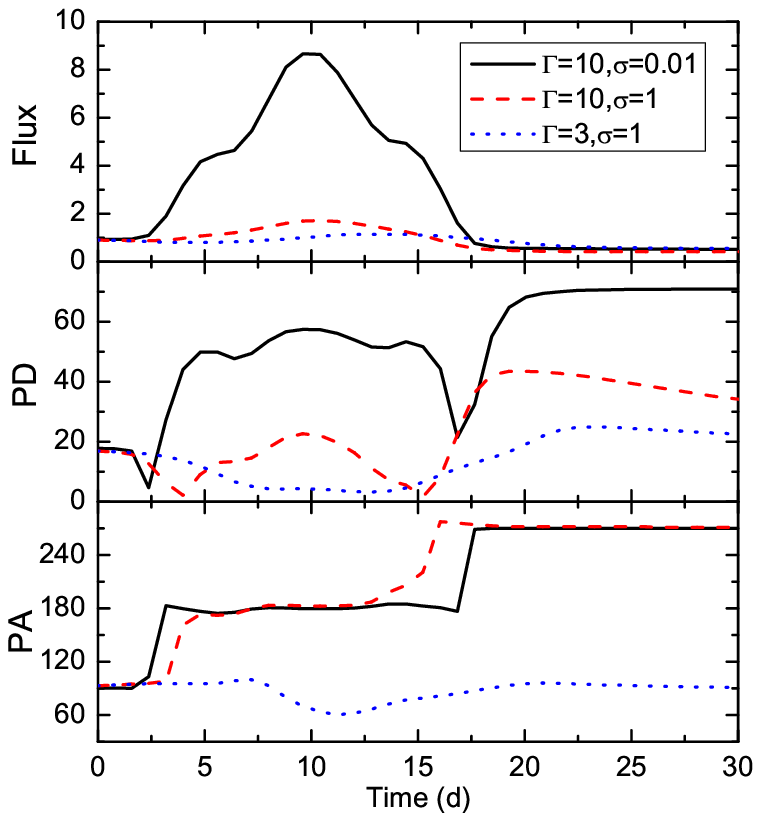}
\caption{Time-dependent radiation and polarization signatures. The x axis is time in days. 
The upper panel shows the relative flux in the optical band. The middle panel is the optical 
PD in percentage. The lower panel is the corresponding PA in degree. Black solid lines are for 
Case I, red dash for Case II, and blue dot for Case III. \label{pol1}}
\end{figure}

In this section, we will investigate how the shock speed and the magnetization in the 
blazar emission environment will affect the time-dependent radiation and polarization 
signatures. Since we generate the shock by a relativistic disturbance, and the simulation
is done in the comoving frame of the emission region, the shock front and its motion can 
hardly be distinguished from the disturbance. Moreover, as we will see in the following 
results, during the propagation, both the disturbance and the shock speeds may vary in 
time. Hence we use the initial Lorentz factor $\Gamma$ of the disturbance rather than 
the actual time-dependent shock speed for the parameter study. For the same reason, the 
magnetization factor $\sigma'$ in the shock frame will vary in time as well; we choose 
the initial $\sigma$ in the emission region instead. This $\sigma$ is also a direct 
indicator of how much the emission region is magnetized at the beginning. By comparing 
the results with the general observational features, we will be able to constrain the 
physics in the emission environment. For this purpose, we will present three cases, 
namely, $\sigma\sim0.01$, $\Gamma=10$ (Case I), $\sigma\sim 1$, $\Gamma=10$ (Case II) 
and $\sigma\sim 1$, $\Gamma=3$ (Case III). The RMHD simulation results are shown in 
Figs. \ref{mhd1}, \ref{mhd2} and \ref{mhd3} for Case I, II and III, respectively, 
along with the associated radiation and polarization signatures in Fig. \ref{pol1}. 
For all the cases studied, we fix the initial nonthermal electron density $n_b$ and 
the characteristic magnetic field strength $B_0$. The magnetization factor is adjusted 
through changing the plasma density. Since we employ a simple electron spectrum, we only 
study the light curves and polarization patterns in the optical band as examples. 
Additionally, as the background nonthermal electron density and the injection rate 
are set as input parameters, we use the relative flux level, where $1$ is approximately 
$10^{-11}~erg~cm^{-2}~s^{-1}$.

\begin{figure}[ht]
\centering
\includegraphics[width=\linewidth]{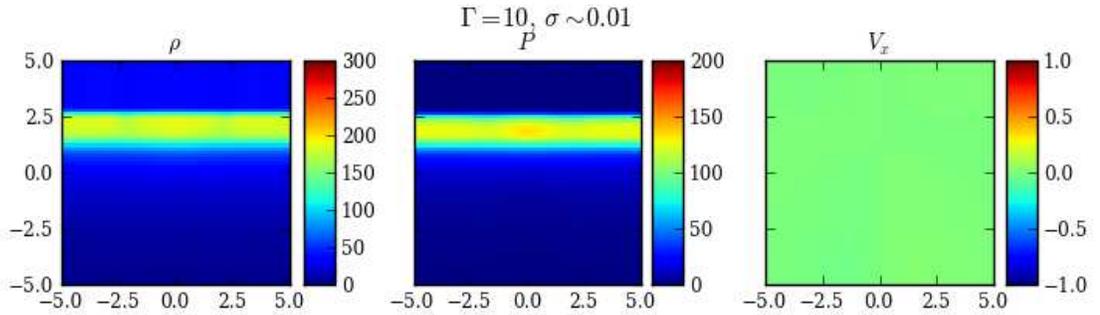}
\caption{Hydro parameters in the $xz$ plane for Case I, when the disturbance is moving out of the emission region. The three panels are the plasma density $\rho$, the pressure $P$, and velocity in $x$ direction $V_x$, which represents the velocity in radial. All quantities are in code units. \label{special}}
\end{figure}

\subsection{Effects of Magnetization}

We first examine a kinetic energy dominated environment, Case I. Such an environment is 
frequently assumed in a shock-initiated flaring model. We first notice that in a trivially 
magnetized setting ($\sigma\sim0.01$), the speed of Alfv\'en waves, $V_A\sim0.1c$, is much 
slower than that of the shock/disturbance. This speed is even slower inside the disturbance. 
As a result, the entire simulation domain can be treated similar to a hydro setup. Consequently, 
we can expect that the shock/disturbance will strongly compress the plasma at the shock front, 
and push it along with the propagation. However, in the ideal MHD, the plasma and the field lines 
are coupled (frozen-in field lines). Therefore, the toroidal magnetic field lines at the shock 
front will be greatly compressed, leading to a much larger toroidal contribution; meanwhile, 
far downstream of the shock, the toroidal lines will be sufficiently stretched, giving rise 
to a perfectly ordered poloidal magnetic field (see Fig. \ref{mhd1}, $B_y$). The physical 
effects described above are the consequence of a shock in a nearly hydro environment with 
the assumption of the frozen-in field lines. Therefore, we expect that the above shock 
modifications to the magnetic field structure is not limited to our initial magnetic field 
topology, but will happen in other magnetic field setups as well. Notice that, however, due to the shock compression, the initial force balance in the radial direction between the magnetic field and plasma pressure and density may be broken. This will lead to some hydrodynamical perturbations in the plasma density and pressure in the radial direction, which are explicitly illustrated in Fig. \ref{special}. However, those perturbations are only transported at sonic speeds. This is supported by Fig. \ref{special} where any plasma velocities in the radial direction that arise from those perturbations are non-relativistic. Therefore, although given enough time they may significantly modify the post-shock structures, they are much slower than the time scale that we are interested in in our RMHD simulation. Therefore, we find that those hydrodynamical effects are of minor importance to our current study, and will not discuss them in the following.

Case II has a moderately magnetized environment ($\sigma\sim 1$). Hence, the magnetic field 
will actively participate in the shock propagation. \cite{Komissarov11} have calculated that 
the maximal relativistic shock compression ratio in a magnetized environment is given by
\begin{equation}
r_c=\frac{6(1+\sigma')}{1+2\sigma'+\sqrt{16\sigma'^2+16\sigma'+1}}
\end{equation}
where $r_c$ and $\sigma'$ are the shock compression ratio and the magnetization factor in 
the shock frame, respectively. As a result, at the shock front, we can see from the simulation 
that the shock compression becomes weaker due to the higher magnetization (Fig. \ref{mhd2}, $B_y$). 
Furthermore, although initially the emission region has a negligible magnetic force, the enhancement 
in the toroidal component at the shock front will break down this balance, resulting in a contracting 
magnetic force in the $r$ direction, given by
\begin{equation}
F_B=-\frac{\partial}{\partial r}\frac{B_{\phi}^2+B_z^2}{2}-\frac{B_{\phi}^2}{r}
\end{equation}
This contraction is transported by Alfv\'en waves, which are mildly relativistic in this case. 
Therefore, in addition to the compression in the $z$ direction at the shock front, there exists 
a contraction of the plasma in the $r$ direction. Because of the frozen-in field lines, the 
poloidal component will be strengthened at the shock front (Fig. \ref{mhd2}, $B_z$). We can 
see from the simulation that at a later stage, the increase in the poloidal component gradually 
gets closer to that in the toroidal component, and even the shape of the shock/disturbance is 
modified, showing a bullet shape at the central part (Fig. \ref{mhd2} $V_z$). Moreover, at 
variance with Case I, far downstream of the shock/disturbance, even though some plasma is 
pushed away by the shock/disturbance, the magnetic field is observed acting to revert to its 
initial topology (Fig. \ref{mhd2}, $B_y$ and $B_z$).

In the spectral variability fitting, it is frequently argued that during the shock propagation, 
the local magnetic field either generally maintains its strength and topology \citep{Sokolov04,Joshi07}, 
or reverts to its initial state after the shock \citep{Chen14,Zhang14}. This requirement is often 
necessary to produce the best fittings. Our simulations demonstrate that this can only happen in an 
adequately magnetized emission region. Therefore, detailed analysis of the magnetic field evolution 
and shock dynamics should play an essential part in those spectral variability models.

We notice that in both simulations, there are regions with negative $V_z$ (Figs. \ref{mhd1} and 
\ref{mhd2}, $V_z$). This is likely due to the reverse shock. In reality,  the reverse shock may 
also lead to some radiation and polarization signatures, but its effect is expected to be weaker 
than the main shock. Thus in the following we will not consider it.

Now we will consider the radiation and polarization signatures resulting from the above RMHD 
simulations, calculated using the 3DPol code. The general trends are similar for both cases. Before 
the shock moves in, the entire emission region is axisymmetric. Although the poloidal and toroidal 
components are comparable in the force-free setup, due to the LOS effect, the projected toroidal 
component onto the plane of sky is weaker than the projected poloidal component. Therefore 
initially the emission region has a total polarization dominated by the poloidal component 
(Fig. \ref{pol1}). When the shock moves in, the toroidal component is strongly increased and 
fresh nonthermal electrons are injected at the shock front. However, due to the LTTEs, only 
a small elliptical region on the right that is near the observer is seen (the near side, see 
Fig. \ref{sketch} (right), red). Therefore, the flux is seen to gradually increase. Nevertheless, 
since the toroidal enhancement is strong, a small flaring region is adequate to push the 
polarization to be dominated by the toroidal component. As a result, the PA quickly rotates 
to $180^{\circ}$ and forms a plateau, which corresponds to the toroidal domination. The PD 
first experiences a drop which indicates a switch in the domination between the two components, 
then climbs up to a high level, implying a strong toroidal dominance (Fig. \ref{pol1}). When 
the shock is about to leave the emission region, the flaring region reaches maximum (Fig. 
\ref{sketch} (right), green). Hence the flux peaks. After that, the flaring region moves far 
from the observer to the left (the far side, Fig. \ref{sketch} (right), blue), so the flux 
gradually decreases in an apparently symmetric pattern. However, as some plasma and the coupled 
magnetic field lines have been pushed away by the shock/disturbance, the total magnetic field 
strength is smaller than the initial state. Hence the flux level is lower than the initial 
value. For the same reason, the toroidal contribution is weakened near the end of the flare, 
hence the PD rises up higher than the initial value, revealing a stronger poloidal contribution. 
Since the helical magnetic field direction on the far side is opposite to that on the near side, 
the PA instead completes a $180^{\circ}$ rotation to the initial value ($180^{\circ}$ PA ambiguity, 
Fig. \ref{pol1}).

The major differences between Case I and II are the following. First, due to the stronger magnetic 
field, Case II has a weaker compression in the toroidal component at the shock front than Case I, 
while it experiences a rise in the poloidal contribution. Therefore, the PA rotation in Case II 
appears smoother and has a shorter toroidal dominated plateau than Case I. Also the maximal flare 
level in Case II is lower. More importantly, near the end of the flare, the moderate magnetization 
in Case II is acting to restore the initial magnetic topology, hence even if the PD rises above 
the initial value, it immediately starts to decrease, indicating the recovery of the toroidal component 
(Fig. \ref{pol1}). \cite{Abdo10} have shown that at the end of a flare + PA swing event in 3C~279, the 
PD rises up above the initial value, which is immediately followed by a significant restoration to 
approximately the initial PD level. The same restoring phases are seen in a number of  flare + PA 
rotation events as well \citep{Larionov08,Larionov13,Morozova14}. On the other hand, in Case I the 
magnetization is too weak, thus the PD rises up to $\sim 70 \%$ at the end of the flare and shows no 
restoration, indicating a purely poloidal magnetic structure. As is mentioned above, such a phenomenon 
is generally the consequence of a trivially magnetized environment, and it is expected to happen with 
other initial magnetic field topologies. Therefore, we argue that such drastic changes in the polarization 
signatures are unrealistic compared to observations, so that a nearly unmagnetized emission environment 
model can be ruled out.

\subsection{Effects of Shock Speed}

The effects of the shock speed are rather straightforward. We will illustrate those by examining 
Case III, which shares the same magnetization as Case II but has a slower disturbance. A slower 
disturbance has less kinetic energy. Therefore, as the shock/disturbance propagates, it cannot 
exert a very strong pressure onto the plasma; in return the plasma will decelerate the shock/disturbance. 
Thus we see in Fig. \ref{mhd3} that the shock/disturbance is considerably slower than the initial value 
($\Gamma=3$). Therefore, the flare duration is much longer than the previous cases (Fig. \ref{pol1}). 
Also the shock/disturbance will provide less compression in the plasma at the shock front. As a result, 
we observe in the simulation that the toroidal component is only moderately increased and stretched 
at the shock front and downstream of the shock/disturbance, respectively (Fig. \ref{mhd3}, $B_y$). 
Moreover, the shower shock/disturbance takes longer time to propagate through the emission region, 
allowing for further information exchange by Alfv\'en waves. Therefore, the contraction of the plasma 
at the shock front, due to the enhanced toroidal component, is seen to catch up with the shock 
compression. This leads to a considerably strengthened poloidal component at the shock front. 
Consequently, the PA rotation disappears; instead it shows fluctuations around the initial value. 
However, at the flare top, the poloidal contribution is only marginally stronger than the toroidal 
contribution. Therefore, the PD still displays a decrease at the middle of the flare (Fig. \ref{pol1}). 
Meanwhile, downstream of the shock/disturbance, Alfv\'en waves restore the initial magnetic field 
topology to a large extent (Fig. \ref{mhd3}, $B_y$ and $B_z$). Hence the final PD is not too much 
higher than the initial value, and gradually recovers after the flare (Fig. \ref{pol1}). Another 
interesting effect is that due to the weaker kinetic energy in the shock/disturbance and longer 
influence of Alfv\'en waves, the shape of the shock/disturbance is distorted (Fig. \ref{mhd3}, 
$V_z$). Thus the polarization patterns, especially the PA, appear less symmetric in time than 
the previous cases (Fig. \ref{pol1}).

\section{Parameter Study}

In this section, we will perform parameter studies to constrain the shock speed and the magnetization 
of the emission environment, by comparing the polarization signatures with the general observational 
properties. Since the injection rate at the shock front is somewhat arbitrary, we will not compare 
the light curves. Although in principle the injection rate will affect the polarization signatures, 
in the following we will demonstrate that the polarization signatures are mainly the consequence of 
the magnetic field evolution. Here we use moderate ($\Gamma=6$) and slow ($\Gamma=3$) speeds, as well 
as weak ($\sigma\sim 0.1$), strong ($\sigma\sim 10$) and moderate ($\sigma\sim 1$) magnetization 
factors as examples.

\subsection{Weakly Magnetized Environment, $\sigma\sim 0.1$}

\begin{figure}[ht]
\centering
\includegraphics[width=\linewidth]{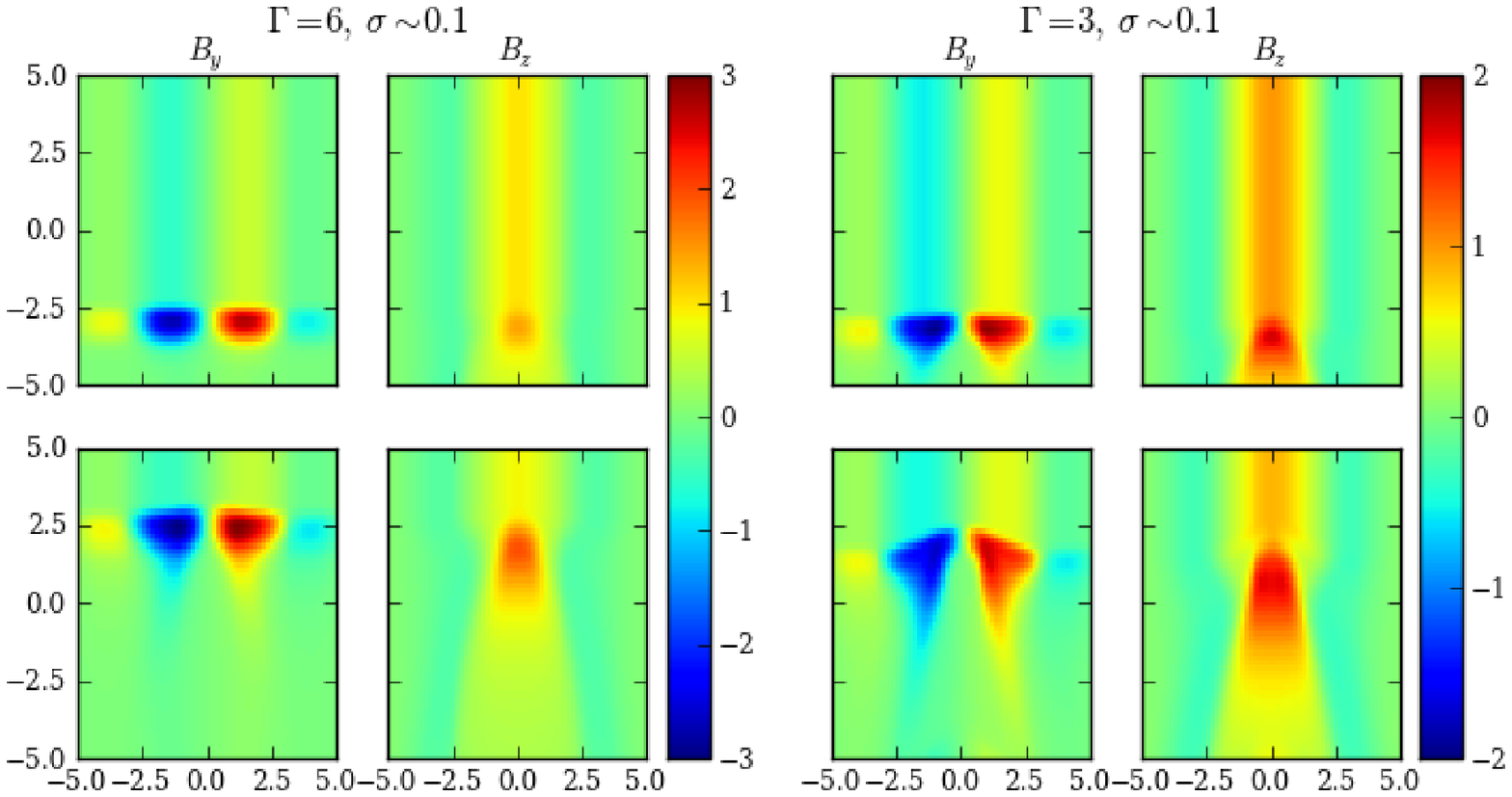}
\caption{RMHD simulation results for a weakly magnetized environment ($\sigma \sim 0.1$). We no 
longer plot the disturbance velocity $V_z$. The left panel is for a medium speed disturbance 
($\Gamma=6$), the right a slow speed disturbance ($\Gamma=3$). Otherwise plots are as in Fig. 
\ref{mhd1}. \label{mhd4}}
\end{figure}

\begin{figure}[ht]
\centering
\includegraphics[width=\linewidth]{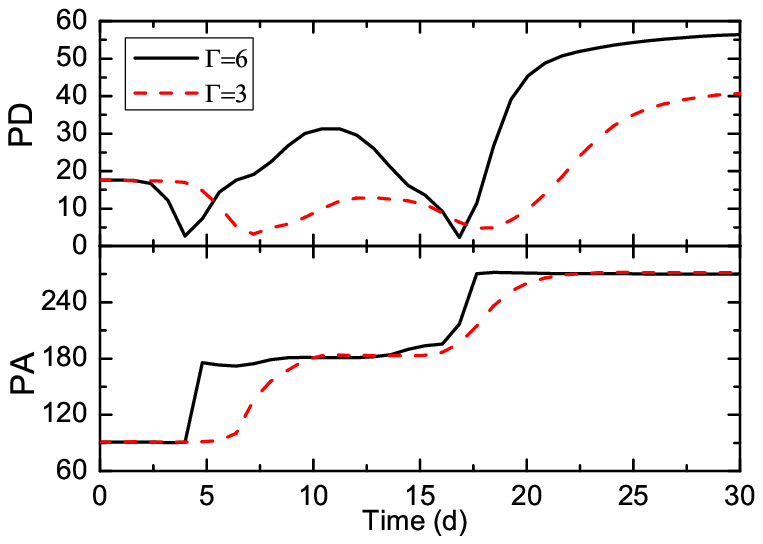}
\caption{Time-dependent radiation and polarization signatures for $\sigma\sim 0.1$. The relative 
flux is not plotted. Black solid lines are for $\Gamma=6$, red dash are for $\Gamma=3$. Otherwise 
panels and curves are as in Fig. \ref{pol1}. \label{pol2}}
\end{figure}

We study a weakly magnetized setup ($\sigma\sim 0.1$), with the application of moderate ($\Gamma=6$) 
and slow ($\Gamma=3$) disturbances. Due to the small magnetization factor, the mediation by Alfv\'en 
waves is slow. Nevertheless, especially in the case of a slow disturbance, the effects of the 
magnetization cannot be overlooked: the poloidal component is slightly enhanced at the shock front, 
and the shape of the shock/disturbance is changed (Fig. \ref{mhd4}, $B_z$ and $V_z$). As a consequence, 
the polarization patterns appear relatively smooth, in particular the PA rotation (Fig. \ref{pol2}). 
However, downstream of the shock/disturbance, Alfv\'en waves fail to restore the initial magnetic 
topology. Hence at the end of the flare, the PD rises up and maintains a high level ($\sim 40\%$), 
without showing any restoring phase (Fig. \ref{pol2}). This implies that in a weakly magnetized 
emission environment, shocks can easily alter the magnetic topology, causing significant changes 
in the polarization signatures. On the contrary, such polarization variations are seldom reported 
in observations. Therefore, we do not favor a weakly magnetized emission environment.

\subsection{Strongly Magnetized Environment, $\sigma\sim 10$}

\begin{figure}[ht]
\centering
\includegraphics[width=\linewidth]{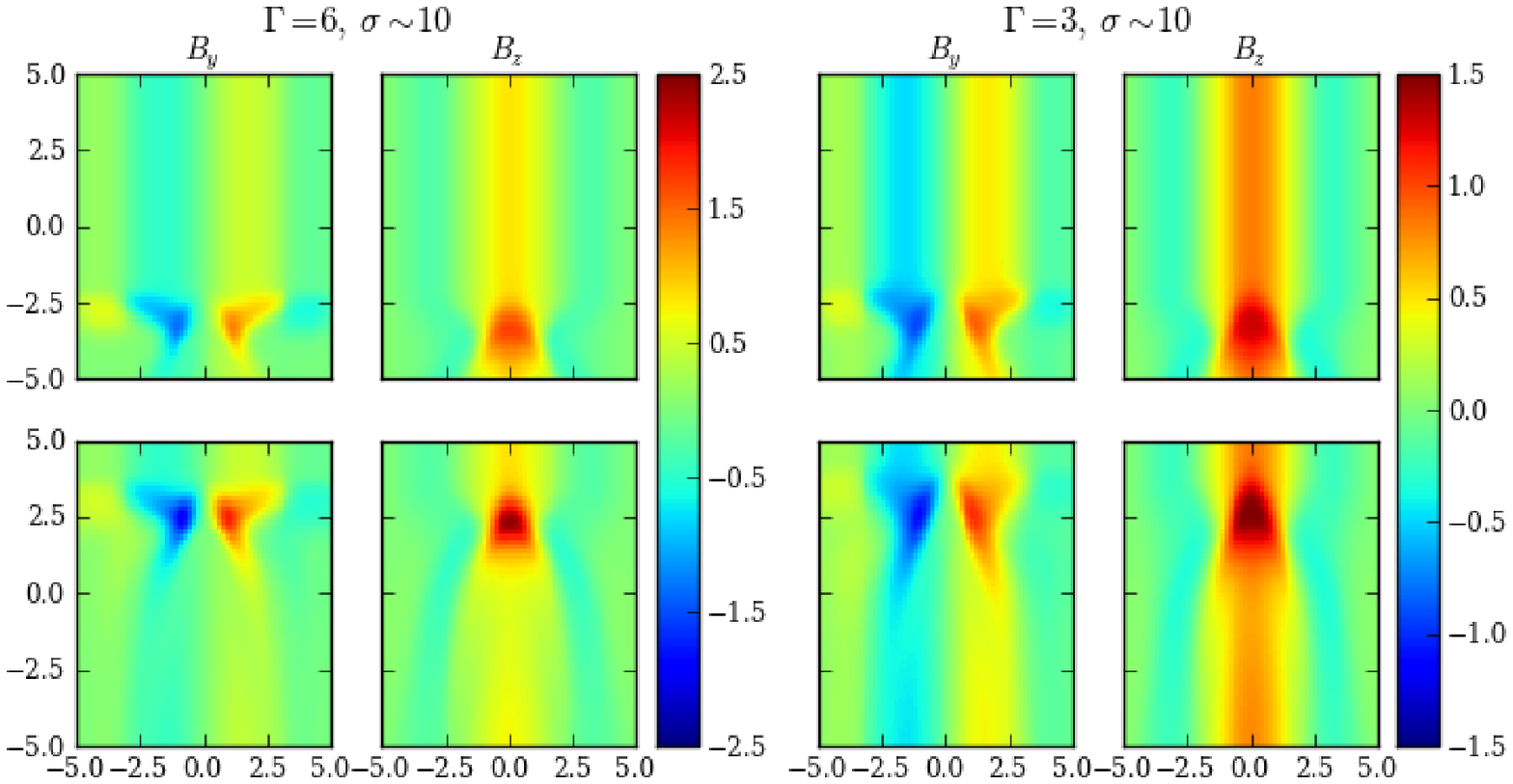}
\caption{RMHD simulation results for a strongly magnetized environment ($\sigma \sim 10$). 
Plots are as in Fig. \ref{mhd4}. \label{mhd5}}
\end{figure}

\begin{figure}[ht]
\centering
\includegraphics[width=\linewidth]{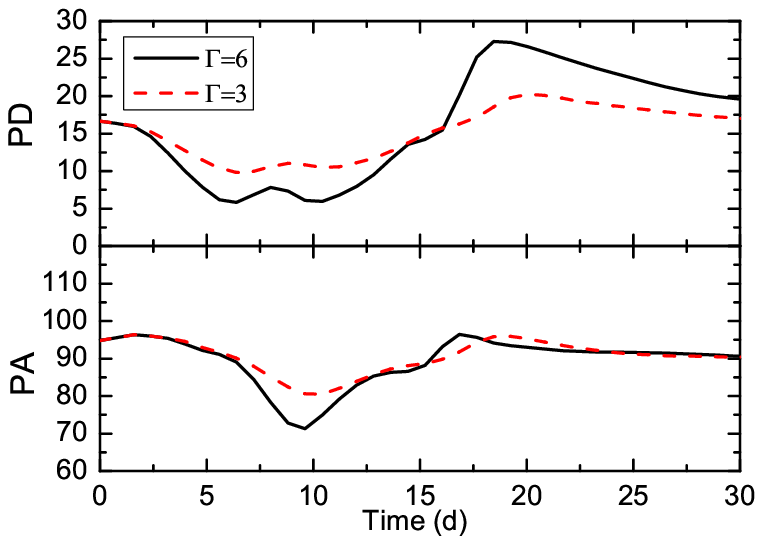}
\caption{Time-dependent radiation and polarization signatures for $\sigma \sim 10$. Panels and 
curves are as in Fig. \ref{pol2}. \label{pol3}}
\end{figure}

Next we consider an emission region with a high magnetization factor ($\sigma\sim 10$). Here 
Alfv\'en waves become relativistic, thus the shock compression is much weaker. All effects of 
the magnetization as in Case II and III show up. The difference is that even in the case of a 
medium speed disturbance ($\Gamma=6$), the increase in the poloidal component is higher than 
that in the toroidal field at the shock front (Fig. \ref{mhd5}, $B_y$ and $B_z$). Therefore, 
the PA patterns for both cases exhibit no rotation (Fig. \ref{pol3}). Additionally, the 
shock/disturbance is greatly distorted. Hence both the PD and the PA patterns appear rather 
erratic, showing a lot of bumps throughout the flare (Fig. \ref{pol3}), even though the general 
geometry is rather symmetric. Finally, the restoration downstream of the shock/disturbance is 
very fast (Fig. \ref{mhd5} $B_y$ and $B_z$), thus the PD only rises up gently above the initial 
value, and quickly recovers at the end of the flare (Fig. \ref{pol3}).

Notice that for a sufficiently fast shock/disturbance, a PA rotation may reappear. However, we 
expect that such a rotation is likely to be very smooth, without a long toroidal dominated 
plateau as in Case I to III. Meanwhile the PD variation can be within some small value. This 
may agree with the observed flare + PA rotation events. To summarize, a strongly magnetized 
emission environment may generate the typical erratic perturbations in the polarization 
signatures in the condition of a relatively slow disturbance, together with the smooth 
flare + PA rotation events provided a relatively fast disturbance. Thus we prefer an emission 
region with high magnetization.

\subsection{Transition Point: Moderately Magnetized Environment, $\sigma\sim 1$}

\begin{figure}[ht]
\centering
\includegraphics[width=0.49\linewidth]{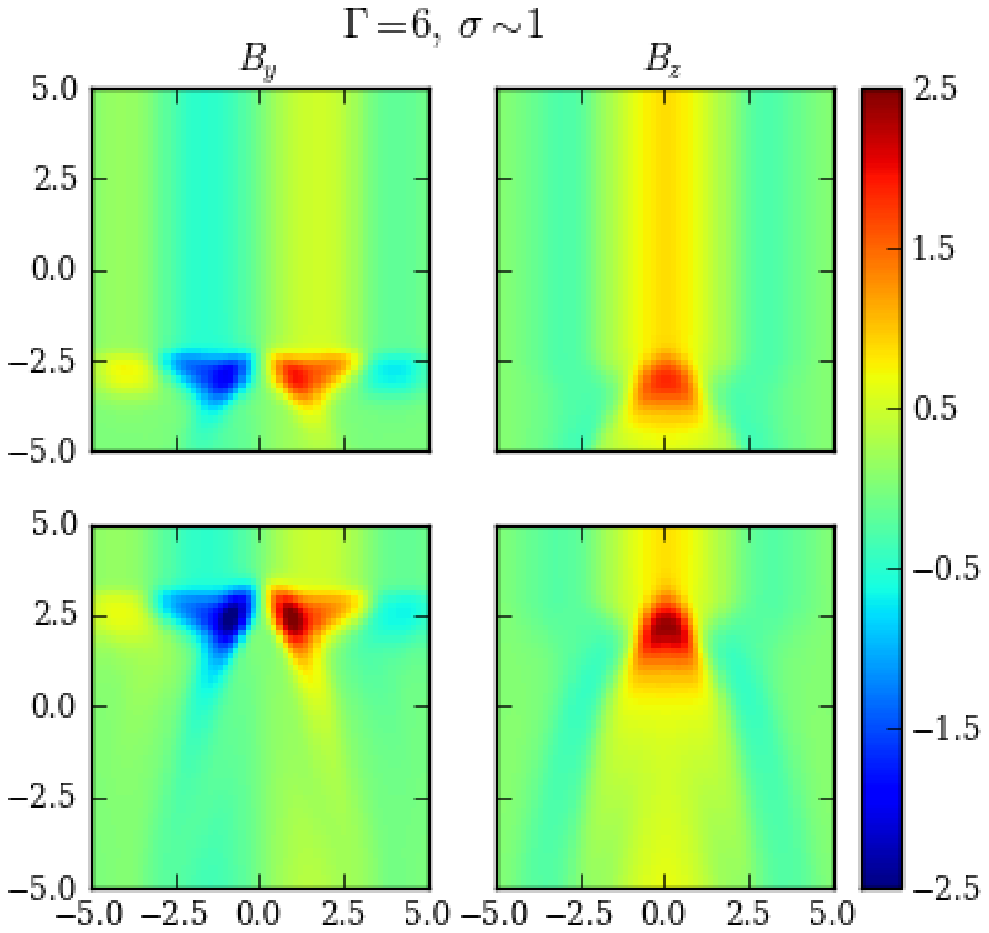}
\includegraphics[width=0.49\linewidth]{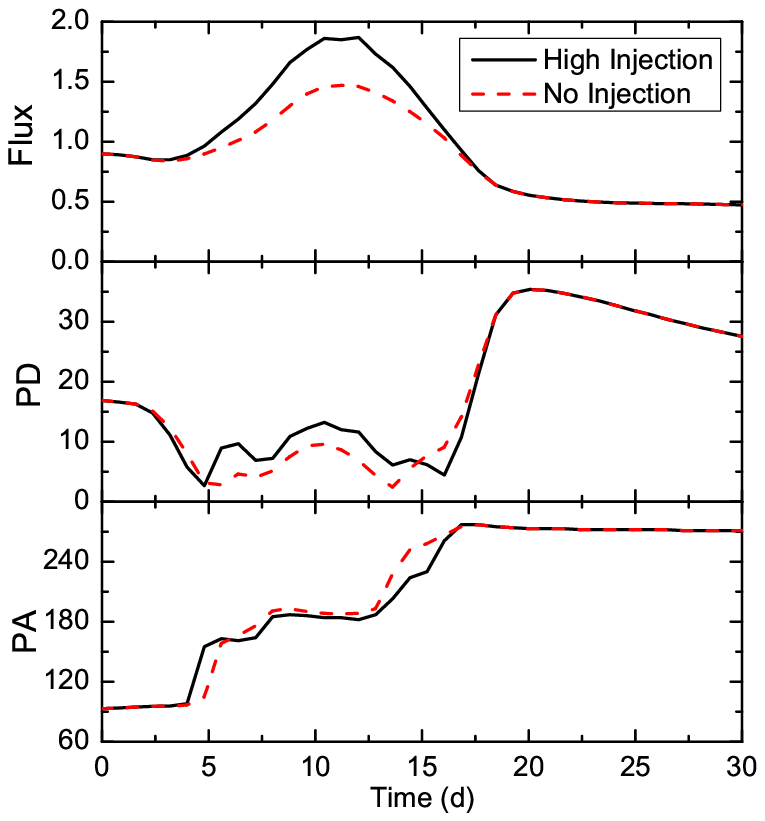}
\caption{RMHD simulations and the associated time-dependent radiation and polarization signatures for 
$\Gamma=6$, $\sigma\sim 0.1$. The left side is the RMHD simulation results; plots are similar in Fig. 
\ref{mhd4}. The right side is the 3DPol results. Black solid lines are for the high injection case, 
red dash are for the no injection case; otherwise panels and curves are as in Fig. \ref{pol1}. 
\label{mhdpol}}
\end{figure}

Finally we take a look at an intermediate condition, $\Gamma=6$ and $\sigma \sim 1$. This case 
shares the same magnetization factor as Case II and III. From the simulation we can see that 
the general features are similar, but the increases in the poloidal and the toroidal components 
at the shock front are nearly identical (Fig. \ref{mhdpol} (left), $B_y$ and $B_z$). However, 
since the emission region is a cylinder, a larger fraction (at larger radii) of the emission 
region possesses the enhanced toroidal component. Hence we expect that the radiation during the
flare should have more toroidal contributions. In order to test the effects of the injection 
rate, we perform two 3DPol runs: one with an artificially increased injection rate (approximately 
ten times higher than the normal value), and one with no injection; the results are shown in 
Fig. \ref{mhdpol} (right). We observe that although the flare level of the high injection case 
is much higher than the no injection case, the polarization variations appear similar (Fig. 
\ref{mhdpol} (right)). This indicates that the polarization signatures in our calculation are 
largely due to the magnetic field evolution, instead of the ad hoc nonthermal electron injection 
rate. We can see that the PA rotation does not show significant toroidal dominated plateau. A 
number of flare + PA rotation events feature continuous PA rotations without any obvious plateau 
step \citep[e.g.,][]{Marscher08,Abdo10,Chandra15}. Based on our simulations, we suggest that 
such events are likely originating from a sufficiently magnetized environment with a relatively 
fast disturbance traveling through. Also in this case a vigorous restoring phase is present.

\section{Discussions and Summary}

We have presented the first polarization-dependent radiation modeling based on RMHD simulations of 
relativistic shocks in helical magnetic fields in the blazar emission region. We find in 
Section 3 that the magnetic field evolution during the shock is intrinsically governed by a 
competition between the shock speed and the magnetization in the emission region: the shock/disturbance 
tries to alter the magnetic topology, while the magnetization attempts to resist any modifications. 
We then investigate the parameter space in Section 4 to further constrain the shock parameters and 
the magnetization in the emission environment. In the following we will summarize the major features 
and illustrate their connections to observations.

We set up an emission region initially pervaded by a purely helical magnetic field with comparable 
poloidal and toroidal contributions. Due to the LOS effect, at the beginning the polarization is 
dominated by the poloidal component. A flat relativistic disturbance, traveling parallel to $B_z$, 
will then form a shock wave and propagate through the emission region. In a weakly magnetized 
environment ($\sigma < 1$), the magnetic field cannot compete against the shock/disturbance. 
Hence even a mildly relativistic shock/disturbance can permanently change the magnetic topology 
to a large extent (here "permanently" means on a time scale that is much longer than the flare 
duration). This results in drastic variations in the polarization signatures, with no signs of 
recovery. Such phenomena are hardly detected, thus we conclude that a weakly magnetized emission 
region is not favored.

In a strongly magnetized emission region ($\sigma \gtrsim 1$), the magnetic field topology is 
able to survive the impact of the shock/disturbance. For the slow cases, the shock/disturbance 
is unable to build up a strong toroidal component, thus the polarization signatures usually 
exhibit rather erratic fluctuations, which are typical in observations. If the shock/disturbance 
is slightly faster, then it is plausible to build up the toroidal component and lead to significant 
polarization variations such as the PA rotations; but in both cases, the flux will not increase 
sufficiently to produce a flare. Such polarization variations with no obvious flares are consistent 
with some observations \citep[e.g.,][]{Itoh13,Jorstad13}. However, \cite{Kirk00} and \cite{Achterberg01} 
have shown that for relativistic shocks, the hardest obtainable particle power-law index is 
approximately
\begin{equation}
p=\frac{r_c+2}{r_c-1}
\end{equation}
where $p$ is the power-law index and $r_c$ is the shock compression ratio mentioned previously. We 
estimate that the strongest shock in our slow disturbance cases can only reach a nonthermal particle 
spectrum of power-law index of $\sim 4$, which is too soft to fit the general blazar SEDs. In addition, 
relativistic collisionless shocks in the highly magnetized regime may be unable to generate fluctuating 
magnetic fields and hence inhibit Fermi acceleration \citep[e.g.,][]{Sironi15}. Therefore, a slow 
shock/disturbance may not efficiently accelerate nonthermal particles that are necessary to generate 
a flare. This poses a tricky problem for slow shocks. On the other hand, for a relatively fast 
shock/disturbance, the polarization variations can either fluctuate around some mean value or 
smoothly rotate in the PA and present a restoring phase at the end of the flare; both are detected 
in observations \citep[e.g.,][]{Larionov08,Abdo10,Blinov15}. In summary, we prefer a strongly 
magnetized emission region, in which fast disturbances can be the driver of the flaring activities.

Observations and spectral variability fittings have shown that the typical Lorentz factor of the 
blazar emission region is within a few tens \citep[e.g.,][]{Jorstad05,Boettcher13}. This is likely 
to be the upper limit of the shock speed. Therefore, a sufficiently magnetized emission environment 
will mostly prohibit any dramatic polarization variations, such as the PA rotations and the sudden 
changes in the PD (flares in the polarized flux). For the blazar emission regions with a weaker 
magnetization, the PD flares and the PA rotations are triggered by strong shocks. Hence we expect 
strong nonthermal electron injections. Therefore, intense polarization variations are probably 
accompanied by vigorous multiwavelength flares. These findings are consistent with observations 
\citep{Blinov15}. Another implication is that the blazar emission region presumably maintains a 
relatively high magnetization, thus it cannot dissipate too much magnetic energy during the 
shock-initiated flaring activities.

An alternative may weaken some of the constraints mentioned above. In a highly magnetized region, 
magnetic reconnection can also be the driver of flares. Several authors have illustrated that 
reconnection can efficiently accelerate nonthermal particles to form a power-law spectrum 
\citep{Guo14,Sironi14,Li15}. In our simulations, no preexisting turbulent magnetic field is 
present; but in reality, some amount of turbulence may exist. Therefore, during the shock 
propagation, magnetic reconnection can happen simultaneously, especially in the compression 
regions at the shock front. This can provide extra nonthermal particles, thus a slow shock in 
a partially turbulent, partially ordered (such as a helical geometry) magnetic field structure 
may generate the necessary nonthermal electrons. Several models investigating the reconnection 
in a highly magnetized environment have been proposed to explain very fast blazar events 
\citep{Giannios09,Deng15,Guo15}. Meanwhile, recent simultaneous fittings of blazar SEDs, 
light curves and polarization signatures also favor a reconnection process \citep{Zhang15,Chandra15}. 
During a reconnection event, the magnetic field topology can be greatly modified, which may give 
rise to some interesting polarization signatures as well. Therefore, we expect that even in a highly 
magnetized environment, if the reconnection drives the flaring activities, considerable polarization 
variations are possible. Future observations featuring both the radiation and polarization signatures, 
along with detailed modelings and simulations of the shock and magnetic reconnetion, can possibly 
distinguish the two mechanisms and further constrain the physics of blazar emission regions.

\acknowledgments{We thank the referee for insightful and constructive comments. HZ thanks Fan Guo, 
Shengtai Li and Xiaocan Li for helpful discussions. HZ, WD and HL are supported by the LANL/LDRD 
program and by DoE/Office of Fusion Energy Science through CMSO. MB acknowledges support by the 
South African Research Chairs Initiative (SARChI) of the Department of Science and Technology 
and the National Research Foundation \footnote{Any opinion, finding and conclusion or recommendation 
expressed in this material is that of the authors and the NRF does not accept any liability in this 
regard.} of South Africa. Simulations were conducted on LANL's Institutional Computing machines.}

\clearpage


\begin{thebibliography}{}

\bibitem[Abdo et al.(2010)]{Abdo10} Abdo, A. A., et al., 2010, Nature, 463, 919

\bibitem[Achterberg et al.(2001)]{Achterberg01} Achterberg, A., 
Gallant, Y. A., Kirk, J. G., \& Guthmann, A. W.\ 2001, \mnras, 328, 393

\bibitem[Aharonian et al.(2007)]{Aharonian07} Aharonian, F. A., et al., 2007, ApJ, 664, L71

\bibitem[Albert et al.(2007)]{Albert07} Albert, J., Aliu, E., Anderhub, H., et al.\ 2007, \apj, 669, 862 

\bibitem[Barnacka et al.(2014)]{Barnacka14} Barnacka, A., Moderski, R., Behera, B., Brun, P., \& Wagner, S.,
2014, A\&A, 567, A113

\bibitem[Blinov et al.(2015)]{Blinov15} Blinov, D., Pavlidou, 
V., Papadakis, I., et al.\ 2015, arXiv:1505.07467 

\bibitem[B\"ottcher \& Dermer(2010)]{Boettcher10} B\"ottcher, M., \& Dermer, C. D., 2010, ApJ, 711, 445

\bibitem[B\"ottcher et al.(2013)]{Boettcher13} B\"ottcher, M.,
Reimer, A., Sweeney, K., \& Prakash, A., 2013, ApJ, 768, 54

\bibitem[Chandra et al.(2015)]{Chandra15} Chandra, S., Zhang, H., 
Kushwaha, P., et al.\ 2015, arXiv:1507.06473 

\bibitem[Chatterjee et al.(2012)]{Chatterjee12} Chatterjee, R., Bailyn, C. D., Bonning, E. W.,
Buxton, M., Coppi, P., Fossati, G., Isler, J., Maraschi, L., \& Urry, C. M., 2012, ApJ, 749, 191

\bibitem[Chen et al.(2014)]{Chen14} Chen, X., Chatterjee, R., 
Zhang, H., et al.\ 2014, \mnras, 441, 2188 

\bibitem[Ciprini(2011)]{Ciprini11} Ciprini, S., 2011, in proc. of 2011 Fermi Symposium --- eConf C110509
(arXiv:1112.2639)

\bibitem[Deng et al.(2015)]{Deng15} Deng, W., Li, H., Zhang, 
B., \& Li, S.\ 2015, \apj, 805, 163 

\bibitem[Dermer et al.(1992)]{Dermer92} Dermer, C. D., et al., 1992, A\&A, 256, L27

\bibitem[Giannios et al.(2009)]{Giannios09} Giannios, D., 
Uzdensky, D.~A., \& Begelman, M.~C.\ 2009, \mnras, 395, L29 

\bibitem[Graff et al.(2008)]{Graff08} Graff, P.~B., 
Georganopoulos, M., Perlman, E.~S., \& Kazanas, D.\ 2008, \apj, 689, 68 

\bibitem[Guo et al.(2014)]{Guo14} Guo, F., Li, H., Daughton, W. \& Liu, Y., 2014, PRL, 113, 155005

\bibitem[Guo et al.(2015)]{Guo15} Guo, F., Liu, Y.-H., 
Daughton, W., \& Li, H.\ 2015, \apj, 806, 167 

\bibitem[Itoh et al.(2013)]{Itoh13} Itoh, R., Fukazawa, Y., 
Tanaka, Y.~T., et al.\ 2013, \apjl, 768, L24 

\bibitem[Joshi \& B\"ottcher(2007)]{Joshi07} Joshi, M. \& B\"ottcher, M., 2007, ApJ, 662, 884

\bibitem[Jorstad et al.(2005)]{Jorstad05} Jorstad, S.~G., 
Marscher, A.~P., Lister, M.~L., et al.\ 2005, \aj, 130, 1418 

\bibitem[Jorstad et al.(2013)]{Jorstad13} Jorstad, S.~G., 
Marscher, A.~P., Smith, P.~S., et al.\ 2013, \apj, 773, 147 

\bibitem[Kiehlmann et al.(2013)]{Kiehlmann13} Kiehlmann, S., 
Savolainen, T., Jorstad, S.~G., et al.\ 2013, European Physical Journal Web 
of Conferences, 61, 06003 

\bibitem[Kirk et al.(2000)]{Kirk00} Kirk, J. G., Guthmann, 
A.~W., Gallant, Y.~A., \& Achterberg, A.\ 2000, \apj, 542, 235

\bibitem[Komissarov \& Lyutikov(2011)]{Komissarov11} Komissarov, S.~S., \& Lyutikov, M.\ 2011, \mnras, 414, 2017

\bibitem[Laing(1980)]{Laing80} Laing, R.~A.\ 1980, \mnras, 193, 439

\bibitem[Laing \& Bridle(2014)]{Laing14} Laing, R.~A., \& Bridle, A.~H.\ 2014, \mnras, 437, 3405 

\bibitem[Larionov et 
al.(2008)]{Larionov08} Larionov, V.~M., Jorstad, S.~G., Marscher, A.~P., et al.\ 2008, \aap, 492, 389 

\bibitem[Larionov et al.(2013)]{Larionov13} Larionov, V. M., et al., 2013, ApJ, 768, 40

\bibitem[Li \& Li(2003)]{Li03} Li, S., \& Li,H., 2003, Los Alamos National Lab. Tech. Rep.
LA-UR-03-8935

\bibitem[Li et al.(2015)]{Li15} Li, X., Guo, F., Li, H., 
\& Li, G.\ 2015, arXiv:1505.02166 

\bibitem[Lyutikov et al.(2005)]{Lyutikov05} Lyutikov, M., Pariev, V. I., \& Gabuzda, D. C., 2005, MNRAS, 360, 869

\bibitem[Mannheim \& Biermann(1992)]{Mannheim92} Mannheim, K., \& Biermann, P. L., 1992, A\&A, 253, L21

\bibitem[Marscher \& Gear(1985)]{Marscher85} Marscher, A. P. \& Gear, W. K., 1985, ApJ, 298, 114

\bibitem[Marscher et al.(2008)]{Marscher08} Marscher, A. P., et al., 2008, Nature, 452, 966

\bibitem[Marscher(2014)]{Marscher14} Marscher, A. P., 2014, ApJ, 780, 87

\bibitem[Morozova et al.(2014)]{Morozova14} Morozova, D. A., et al., AJ, 148, 42

\bibitem[M\"{u}cke \& Protheroe(2001)]{Mucke01} M\"{u}cke, A., \& Protheroe, R.J., 2001, Astropart. Phys, 15, 121

\bibitem[Pushkarev et al.(2005)]{Pushkarev05} Pushkarev, A. B., Gabuzda, D. C., Vetukhnovskaya, Yu. N., \& Yakimov, V. E., 2005, MNRAS, 356, 859

\bibitem[Scarpa 
\& Falomo(1997)]{Scarpa97} Scarpa, R., \& Falomo, R.\ 1997, \aap, 325, 109 

\bibitem[Sikora et al.(1994)]{Sikora94} Sikora, M., et al., 1994, ApJ, 421, 153

\bibitem[Sironi 
\& Spitkovsky(2014)]{Sironi14} Sironi, L., \& Spitkovsky, A.\ 2014, \apjl, 783, L21 

\bibitem[Sironi et al.(2015)]{Sironi15} Sironi, L., Keshet, U., 
\& Lemoine, M.\ 2015, arXiv:1506.02034

\bibitem[Sokolov et al.(2004)]{Sokolov04} Sokolov, A., Marscher, 
A.~P., \& McHardy, I.~M.\ 2004, \apj, 613, 725 

\bibitem[Spada et al.(2001)]{Spada01} Spada, M., Ghisellini, 
G., Lazzati, D., \& Celotti, A.\ 2001, \mnras, 325, 1559 

\bibitem[Spitkovsky(2008)]{Spitkovsky08} Spitkovsky, A.\ 2008, \apjl, 682, L5 

\bibitem[Summerlin \& Baring(2012)]{Summerlin12} Summerlin, E.~J., \& Baring, M.~G.\ 2012, 
\apj, 745, 63 

\bibitem[Tavecchio et al.(2010)]{Tavecchio10} Tavecchio, F., Ghisellini, G., Bonnoli, G., \&
Ghirlanda, G., 2010, MNRAS, 405, L94

\bibitem[Zhang et al.(2014)]{Zhang14} Zhang, H., Chen, X. \& B\"ottcher, M., 2014, ApJ, 789, 66

\bibitem[Zhang et al.(2015)]{Zhang15} Zhang, H., Chen, X., 
B{\"o}ttcher, M., Guo, F., \& Li, H.\ 2015, \apj, 804, 58


\end{thebibliography}
\end{document}